\documentclass[final,5p,times,twocolumn]{article}
\usepackage{natbib}
\usepackage{url}
\usepackage{graphicx}

\begin{document}

\title{Asymmetry in galaxy spin directions: a fully reproducible experiment using HSC data}   

\author{Lior Shamir \\ Kansas State University \\ 1701 Platt St, Manhattan, KS 66506, USA}

\date{}
\maketitle




\begin{abstract}
The asymmetry in the large-scale distribution of the directions towards spiral galaxies rotate has been observed by multiple telescopes, all show a consistent asymmetry in the distribution of galaxy spin directions as observed from Earth. Here, galaxies with redshift from HSC DR3 are annotated by their direction of rotation, and their distribution is analyzed. The results show that galaxies that rotate in the opposite direction relative to the Milky Way as observed from Earth are significantly more prevalent compared to galaxies that rotate in the same direction relative to the Milky Way. The asymmetry also forms a dipole axis that becomes stronger when the redshift gets higher. These results are aligned with observations from virtually all premier digital sky surveys, as well as space telescopes such as HST and JWST. That shows that the distribution of galaxy spin directions as observed from Earth is not symmetric, and has a possible link to the rotational velocity of the Milky Way. The experiment is provided with data, code, and a full protocol that allows to easily reproduce the results in a transparent manner. That practice is used to overcome the ``reproducibility crisis" in science. 
\end{abstract}


\section{Introduction}
\label{introduction}

The distribution of the directions of rotation of spiral galaxies has been a topic of study for several decades, often with conflicting results. While the null hypothesis is that the number of galaxies rotating in one direction is the same as the number of galaxies rotating in the opposite direction, multiple studies starting as early as the 1980s have provided consistent results that the distribution of galaxies in the sky does not necessarily satisfy the null hypothesis. At the same time, other studies argued that the null hypothesis agrees with the observations. 

Early observations were based on a relatively small number of galaxies classified manually by the shape of the galaxy arms, showing asymmetry between the number of spiral galaxies that rotate clockwise and the number of spiral galaxies that rotate counterclockwise with confidence of 92\% \citep{macgillivray1985anisotropy}. The deployment of robotic telescopes allowed using far more galaxies, using autonomous digital sky surveys such as the Sloan Digital Sky Survey (SDSS) \citep{longo2011detection,shamir2012handedness,shamir2019large,shamir2020patterns,shamir2020large,shamir2021particles}.

Other digital sky surveys also showed asymmetry between the number of galaxies rotating clockwise and the number of galaxies rotating counetrclockwise. These experiments include DECam \citep{shamir2021large}, Pan-STARRS \citep{shamir2020patterns}, the Dark Energy Survey \citep{shamir2022asymmetry,shamir2022large}, and the DESI Legacy Survey \citep{shamir2022analysis}. These sky surveys cover both the Northern and Southern hemispheres, and showed that the asymmetry in the distribution of galaxy spin directions forms a cosmological-scale dipole axis \citep{shamir2020patterns,shamir2022asymmetry,shamir2021large,shamir2022analysis}. Other experiments were based on space telescopes such as Hubble Space Telescope \citep{shamir2020pasa}, and James Webb Space Telescope \citep{shamir2024galaxy}. The James Webb Space Telescope deep field image acquired inside the field of HST's Ultra Deep Field allows to notice the asymmetry by simple manual inspection \citep{shamir2024galaxy}.

The asymmetry and the large-scale axis exhibited by it might be related to anomalies in the large-scale structure of the Universe, and can agree with other theories such as rotating Universe \citep{godel1949example,ozsvath1962finite,ozsvath2001approaches,sivaram2012primordial,chechin2016rotation,seshavatharam2020integrated,camp2021}, or black hole cosmology \citep{pathria1972universe,easson2001universe,seshavatharam2014understanding,poplawski2010radial,tatum2018flat,christillin2014machian,seshavatharam2020light,chakrabarty2020toy}. On the other hand, the locations of the most likely axis in all experiments are within close proximity to the Galactic pole \citep{mcadam2023asymmetry}. That leads to the possibility that the dipole axis is not necessarily of cosmological origin, but could be due to differences in the brightness of the galaxies driven by their rotational velocity relative to the Galactic pole \citep{mcadam2023asymmetry,shamir2020asymmetry,shamir2017large,universe10030129}.  

While numerous experiments using several different telescopes showed that the distribution of galaxy spin directions is not fully random, other studied showed no statistically significant asymmetry \citep{iye1991catalog,land2008galaxy,hayes2017nature,iye2020spin,petal}. These claims were addressed in previous studies that aimed at reproduction and careful analysis of these experiment to understand the reason they provided random distribution \citep{shamir2023large,shamir2022analysis,shamir2024galaxy,shamir2022using,shamir2024reproducible}.

This paper shows an experiment using galaxies with redshift imaged by the Hyper Suprime-Cam (HSC). The redshift and the powerful imaging power of HSC allows to analyze galaxies with higher redshift compared to other Earth-based telescopes, and therefore better profile the change of the asymmetry in response to the redshift of the galaxies. Code and data are provided, allowing to easily reproduce the experiment and inspect the consistency of the data. That makes the experiment different from most previous studies of this question. The paper also discusses the question of the distribution of galaxy spin direction in the light of the current ``reproducibility crisis" in science.

\section{Data}
\label{data}

The data used in this experiment is galaxies with redshift imaged by HSC and included in HSC third data release (DR3). HSC uses the powerful 8.2 meter Subaru telescope, providing it with imaging power stronger than other Earth-based digital sky surveys such SDSS, Pan-STARRS, or DES. HSC can therefore provide images with high details of the galaxy shapes, allowing to observe the arms of galaxies at higher redshift ranges compared to any other Earth-based digital sky survey. The downside of HSC is that its footprint size is smaller compared to some other Earth-based digital sky surveys. 

The dataset used in this experiment is the same dataset used in \citep{shamir2024empirical}, and the full details regarding the preparation of the dataset are described in that paper. In summary, the initial set of galaxies included all galaxies with redshift in HSC DR3. Because HSC is not a spectroscopic survey, the redshifts of the galaxies are taken from SDSS DR17. That included 101,415 galaxies with redshift of $0<z<0.3$. 

After the images of all galaxies were downloaded, the galaxies were annotated by their direction of rotation. Annotating 10$^5$ galaxies is definitely a labor intensive task that is largely impractical to do manually. More importantly, manual annotation is subjected to cognitive biases that are very difficult to quantify and control. Therefore, the entire process of annotating the galaxies was performed in a fully automatic manner, and without any human intervention except for inspection of the annotation. The spin directions of galaxies were annotated by the {\it Ganalyzer} algorithm \citep{shamir2011ganalyzer,ganalyzer_ascl} as described in \citep{universe10030129,shamir2024empirical,shamir2021large,shamir2022large,shamir2022analysis,shamir2022analysis2,shamir2022asymmetry}. The {\it Ganalyzer} algorithm and the way it was used is described in a large number of previous papers \citep{shamir2011ganalyzer,universe10030129,shamir2024empirical,shamir2021large,shamir2022large,shamir2022analysis,shamir2022analysis2,shamir2022asymmetry}. 

In summary, {\it Ganalyzer} first converts each galaxy image into its radial intensity plot transformation. Then, for each radial distance from the center of the galaxy, it applies peak detection to identify bright pixels. Because arm pixels are brighter than background pixels, the bright pixels are expected to be on the galaxy arms. If the polar angle of these bright pixels changes with the radial distance, it means that the arm is curved, and the direction of the curve can identify the direction of rotation of the galaxy. That is done by applying a linear regression to the bright pixels on the radial intensity plot, and the sign of the regression coefficient determines the direction of the arm curve. The full details with thorough experimental results are available at \citep{shamir2011ganalyzer,universe10030129,shamir2024empirical,shamir2021large,shamir2022large,shamir2022analysis,shamir2022analysis2,shamir2022asymmetry}.

{\it Ganalyzer} is a fully symmetric algorithm \citep{shamir2011ganalyzer}, and its simple and ``mechanical" nature allows to understand and control the way it works. The explainable nature of {\it Ganalyzer} makes it different from machine learning and deep neural network solutions. The complex nature of these algorithms makes them difficult to fully profile, leading to unexpected biases \citep{ball2023,dhar2022systematic}. 

As described with experimental results in \citep{shamir2021large,shamir2022large,shamir2022analysis,shamir2022analysis2,shamir2022asymmetry}, {\it Ganalyzer} annotates the sign of the direction of rotation of galaxies (clockwise or counterclockwise) with virtually no errors. That is, it would very rarely classify a galaxy that rotates clockwise as a galaxy that rotates counterclockwise. At the same time, the algorithm also rejects galaxies that the algorithm cannot determine their direction of rotation. As described in \citep{shamir2011ganalyzer}, if the absolute value of the coefficient of the linear regression is smaller than 0.35, the slope is determined to be too weak to identify the direction of rotation of the galaxy. In that case, the galaxy is ignored and not used in the analysis. In HSC DR3 image data, $\sim$86\% of the galaxies were ignored, which means that these galaxies were elliptical, irregular, or that their shape did not allow to identify their direction of rotation.

When the annotation of the galaxies was completed, it provided a very clean dataset of 13,477 galaxies \citep{shamir2024empirical}. The dataset can be downloaded at \url{https://people.cs.ksu.edu/~lshamir/data/asymmetry_hsc/}. An inspection of 100 galaxies shows that the annotations of all of these galaxies are in agreement with the annotation of the human eye, and in all cases the annotation is clear and correct. No in-between cases were identified, as such cases are rejected by the algorithms and these galaxies are not assigned with an annotation. The 100 galaxies were selected such that 50 galaxies had redshift lower than 0.1 and 50 had redshift higher than 0.1, to ensure that the correctness of the annotation does not degrade when the galaxies have higher redshifts. The annotation was repeated also after mirroring all galaxy images to ensure that the analysis is symmetric, and that mirroring the galaxy images does not change the annotation. In cases of galaxies with unique shapes or features that can confuse the algorithm, it is expected that such galaxies will be distributed evenly between galaxies that rotate in both directions.

\begin{figure}[h]
\centering
\includegraphics[scale=0.50]{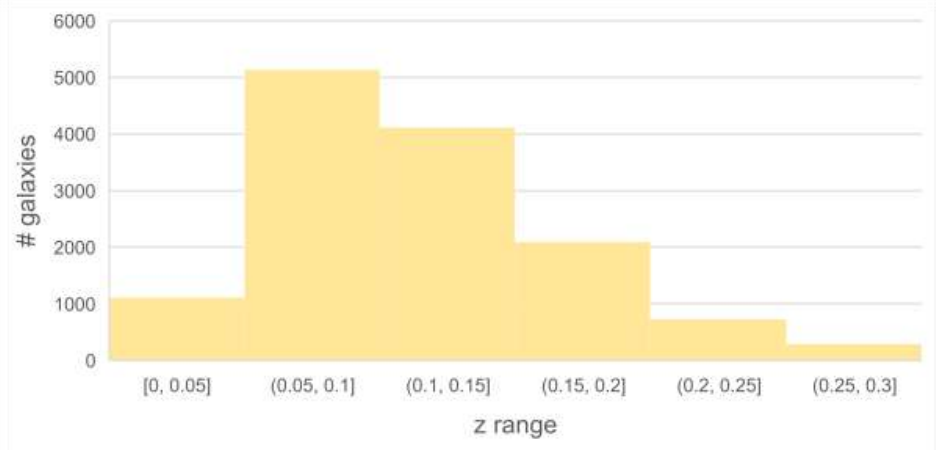}
\caption{The redshift distribution of the galaxies. Since galaxies at higher redshift ranges tend to be smaller and dimmer, the number of annotated galaxies declines as the redshift range gets higher.}
\label{redshift}
\end{figure}

\begin{figure}[h]
\centering
\includegraphics[scale=0.50]{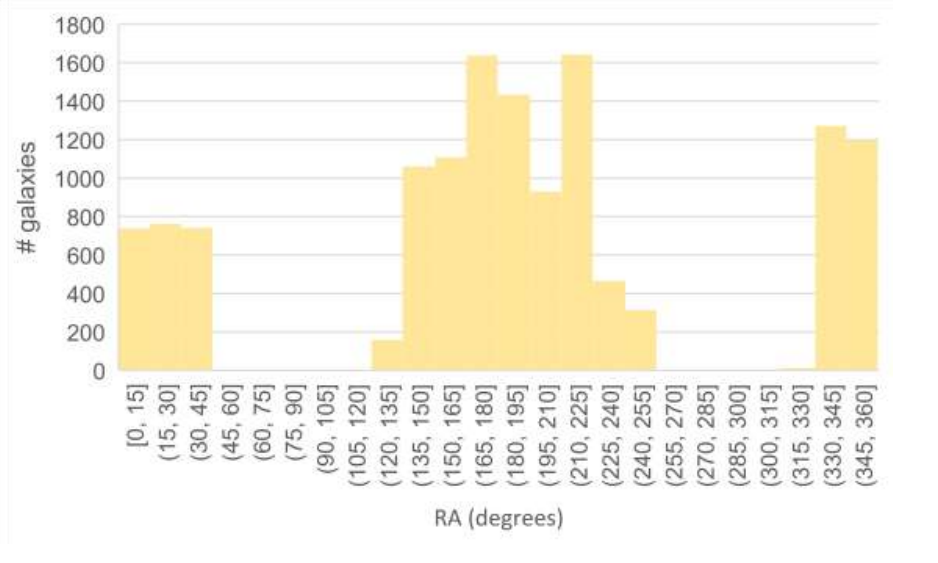}
\caption{RA distribution of the galaxies. The distribution is driven by the footprint of HSC DR3, and therefore some RA ranges do not contain galaxies.}
\label{ra}
\end{figure}

Figure~\ref{redshift} shows the distribution of the redshift of the galaxies. Naturally, the number of annotated galaxies decreases as the redshift gets higher, as at higher redshifts the number of galaxies that their shape can be identified gets smaller. Table~\ref{ra} shows the RA distribution of the galaxies in the dataset, which is driven by the footprint of HSC DR3. Since HSC DR3 does not cover all RA ranges, some RA ranges do not have galaxies in them. The RA ranges of $[45^o,120^o]$ and $[255^o,330^o]$ are not within the HSC DR3 footprint, and therefore contain no galaxies. Since the footprint does not have galaxies in some large parts of the sky, the analysis requires a method that fits the entire set of galaxies into a statistical model, as will be explained in Section~\ref{results}.

\begin{figure}[h]
\centering
\includegraphics[scale=0.50]{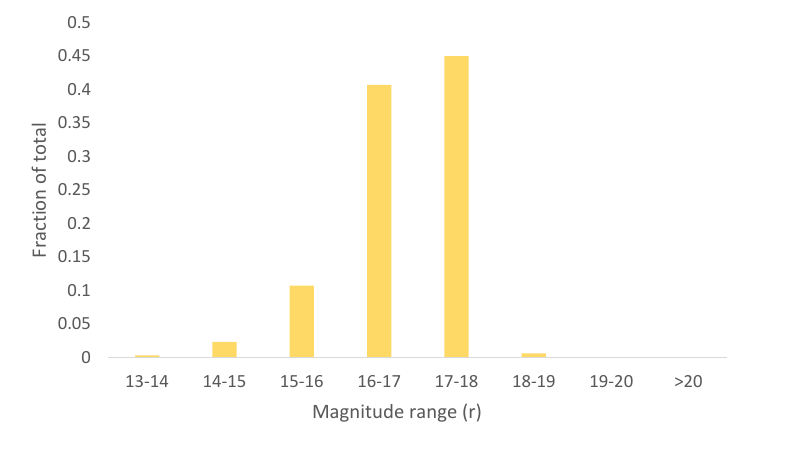}
\caption{The magnitude distribution of the galaxies. The magnitude is the exponential magnitude of the {\it r} band. The vast majority of the galaxies have exponential magnitude of 16-18.}
\label{magnitude}
\end{figure}

Figure~\ref{magnitude} shows the distribution of the exponential magnitude of the galaxies in the {\it r} band. 
As the graph shows, the vast majority of the galaxies are with r magnitude of between 16 to 18, with very rare cases of galaxies dimmer than 18. The dataset also does not contain galaxies brighter than 13.  

\section{Results}
\label{results}

The first experiment was to test whether the distribution of galaxy spin directions is different in the two ends of the Galactic pole. Due to the Doppler shift effect, galaxies that rotate in the opposite direction relative to the Galactic pole are expected to be slightly brighter compared to galaxies that rotate in the same direction relative to the Galactic pole \citep{mcadam2023asymmetry}. With the rotational velocity of the Milky Way of $\sim$ 220 $km \cdot sec^{-1}$, the expected magnitude difference is $\sim0.006$ \citep{mcadam2023asymmetry}. Empirical experiments have indeed shown that the average magnitude of galaxies that rotate in the same direction relative to the Milky Way is different from the average magnitude of galaxies that rotate in the opposite direction. That was observed with data from SDSS, Pan-STARRS \citep{shamir2017large}, Hubble Space Telescope \citep{shamir2020asymmetry} and DESI Legacy Survey \citep{mcadam2023asymmetry}. 

Since galaxies that rotate in the opposite direction relative to the Milky Way are expected to be brighter than galaxies that rotate in the same direction relative to the Milky Way, galaxies that rotate in the same direction relative to the Milky Way are expected to be more prevalent to an Earth-base observed compare to galaxies that rotate in the same direction relative to the Milky Way. Therefore, more galaxies that rotate clockwise are expected to be observed around the Southern Galactic pole, and more galaxies rotating counterclockwise are expected to be observed in the Northern end of the Galactic pole. For instance, JWST deep field images images acquired in close proximity to the Galactic pole showed that the number of galaxies that rotate in the opposite direction relative to the Milky Way is higher than the number of galaxies that rotate in the same direction relative to the Milky Way \citep{shamir2024galaxy}.

HSC DR3 does not cover neither the Southern Galactic pole nor the Northern Galactic pole. But it can be expected that the population of galaxies that are closer to one of the ends of the Galactic poles will exhibit such asymmetry. That means that in the population of galaxies closer to the Northern Galactic pole, galaxies rotating counterclockwise will be brighter, and therefore an excessive number of counterclockwise galaxies will be observed. In the population of galaxies in the Southern end of the Galactic pole, an excessive number of clockwise galaxies is expected. 

Table~\ref{poles} shows the number of galaxies rotating in the same direction relative to the Milky Way and in the opposite direction relative to the Milky Way around the Northern and Southern ends of the Galactic pole. As the table shows, in both ends of the Galactic pole there are more galaxies that rotate in the opposite direction relative to the Milky Way are observed from Earth. As mentioned above, this is not necessarily because galaxies that rotate in the opposite direction relative to the Milky Way are more prevalent in the Universe, but because these galaxies might be slightly brighter due to their rotational velocity relative to the rotational velocity of the Earth within the Milky Way galaxy.

\begin{table}[h]
\centering
\scriptsize
\begin{tabular}{lcccc}
\hline
Pole & \# MW & \# OMW &  $\frac{OMW-MW}{OMW+MW}$  & $p$ value \\
\hline
North  & 4,317   &  4,436  &  0.014  &  0.103 \\   
South  & 2,313   & 2,411  &  0.021  &  0.085 \\  
All    & 6,630   & 6,847     &  0.016  & 0.031  \\
\hline
\end{tabular}
\caption{The number of spiral galaxies that rotate in the same direction relative to the Milky Way and in the opposite direction relative to the Milky Way. The $p$ values are the binomial distribution probability to have such distribution by chance when assuming that the probability of a galaxy to rotate in a certain direction is 0.5}. 
\label{poles}
\end{table}

As the table shows, when using all galaxies in the dataset the probability to have such distribution by mere chance is $p\simeq0.031$. The fact that in both ends of the Galactic pole more observed galaxies rotate in the opposite direction relative to the Milky Way shows the consistency of the asymmetry, but it also shows that it is not a feature of an asymmetry in the galaxy annotation algorithm. Galaxies that rotate in the same direction relative to the Milky Way would seem to rotate counterclockwise in the Southern pole, but clockwise in the Northern Galactic pole. If the algorithm had some unknown bias towards a certain direction, that would have led to more galaxies that rotate in the same direction in one end of the Galactic pole, and a higher number of galaxies that rotate in the opposite direction relative to the Milky Way in the opposite end of the Galactic pole. As mentioned in Section~\ref{data}, the annotation is fully symmetric, and was tested with mirroring the images in this experiment, as well as in numerous previous experiments \citep{shamir2019large,shamir2020patterns,shamir2020pasa,shamir2021particles,shamir2022asymmetry,shamir2022asymmetry,shamir2021large,shamir2022analysis,shamir2024galaxy}.

The separation of the population of galaxies to galaxies that are closer to the Northern Galactic pole and Galaxies that are closer to the Southern Galactic pole provides a very simple experiment that is easy to analyze and reproduce from the data. The statistical inference is also straightforward, and based on simple binomial distribution.

The higher number of galaxies that rotate in the opposite direction relative to the Milky Way is aligned with previous studies. For instance, when using the ``superclean'' annotations of {\it Galaxy Zoo 1} \citep{land2008galaxy} that were also mirrored to offset for the human bias, the asymmetry was 2.1\% when mirroring galaxy images annotated as counterclockwise, and 1.5\% when mirroring galaxy images annotated as clockwise. That is based on the values in Table 2 in \citep{land2008galaxy}, as explained in \citep{shamir2022analysis,mcadam2023reanalysis,shamir2023large,shamir2024galaxy}. The magnitude and direction of the asymmetry agree with other datasets of SDSS galaxies with spectra, showing a difference of $\sim$1.1\% \citep{shamir2020patterns}. The footprint of SDSS galaxies with spectra is such that most galaxies are around the Northern Galactic pole, leading to the asymmetry when using the entire population of galaxies in that footprint \citep{shamir2023large}. 

Another experiment aimed at identifying whether the asymmetry in the distribution of galaxy spin directions exhibits a cosmological-scale dipole axis. That is based on the same analysis done in \citep{shamir2019large,shamir2020patterns,shamir2020pasa,shamir2022asymmetry,shamir2022asymmetry,shamir2021large,shamir2022analysis}, and analyzed thoroughly with extensive tests in \citep{shamir2021particles}. In summary, all possible integer combinations of $(\alpha,\delta)$ are fitted to a cosine dependence with the directions of rotations of the galaxies. That is done by using $\chi^2$ statistics shown in Equation~\ref{chi2}, such that the $\chi^2$ of the observed directions of rotation of the galaxies is compared to the $\chi^2$ when assigning the galaxies with random spin directions. $\phi_i$ is the angular distance between $(\alpha,\delta)$ and galaxy {\it i}, and $d$ is the direction of rotation of the galaxy, where 1 means clockwise and -1 means counterclockwise. 

\begin{equation}
\chi^2_{(\alpha,\delta)}=\Sigma_i | \frac{(d_i \cdot | \cos(\phi_i)| - \cos(\phi_i))^2}{\cos(\phi_i)} |
\label{chi2}
\end{equation}

The $\chi^{2^{\alpha,\delta}}_{real}$ is computed with the observed directions of rotations of the galaxies. The $\chi^2$ when using the random directions of rotations is computed 1,000 times for each integer $(\alpha,\delta)$ combination, and the $\chi^{2^{\alpha,\delta}}_{random}$ is the mean $\overline{\chi^{2_{\alpha,\delta}}}$ of the 1,000 runs where each run has a different set of random directions. The standard deviation $\sigma(\chi^2_{\alpha,\delta})$ is also computed using the 1,000 runs. The $(\alpha,\delta)$ that provides the highest $\frac{\chi^2_{real}-\chi^2_{random}}{\sigma(\chi^2_{\alpha,\delta})}$ is the most likely dipole axis formed by the distribution of the galaxy spin directions. The analysis is explained in full detail in \citep{shamir2019large,shamir2020patterns,shamir2020pasa,shamir2022asymmetry,shamir2022asymmetry,shamir2021large,shamir2021particles,shamir2022analysis}. $\chi^2$ statistics assumes normal distribution of the galaxy spin directions, which is satisfied under the assumption that the sign of the galaxy spin direction and the inclination of the galaxies are expected to be random.

\begin{figure*}[h]
\centering
\includegraphics[scale=0.25]{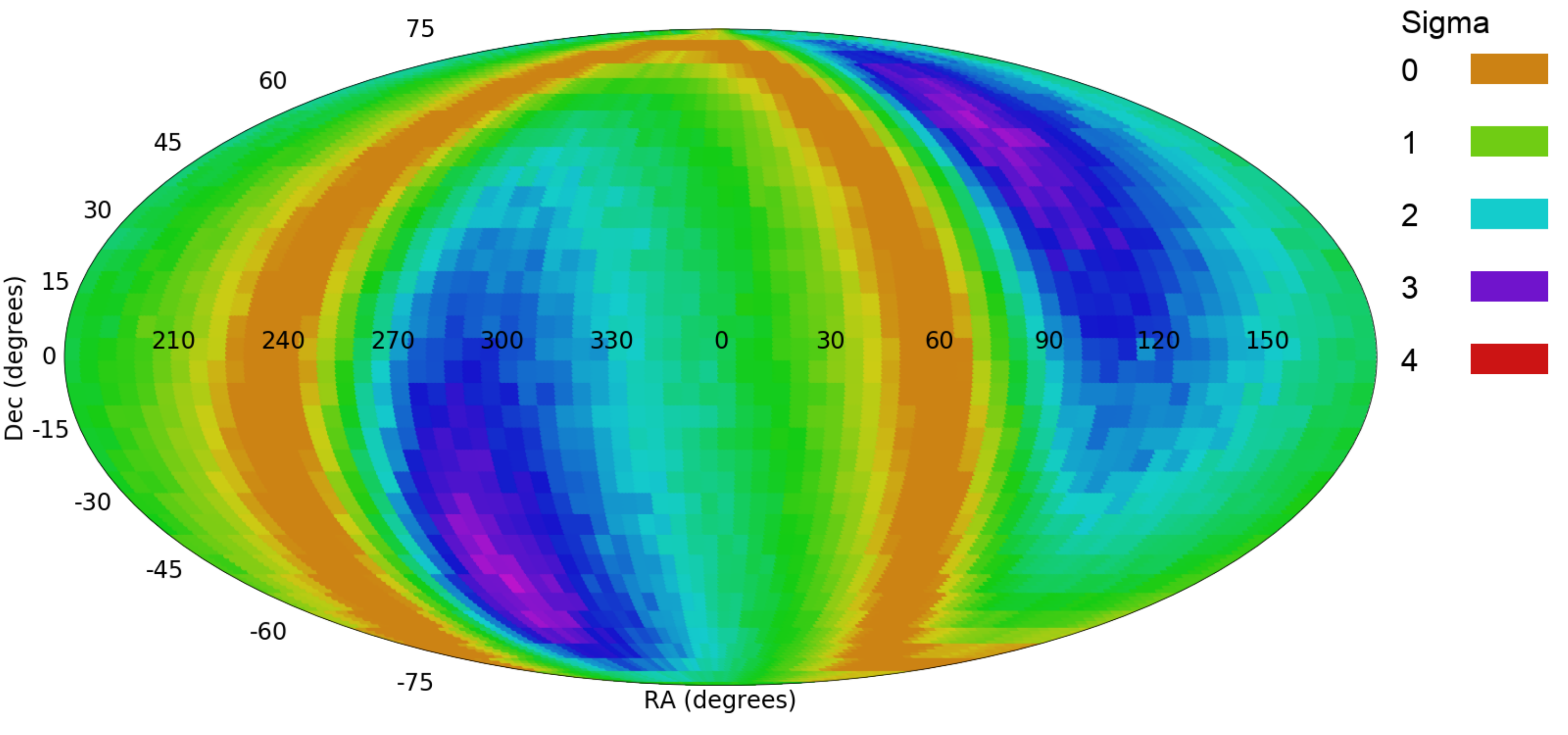}
\caption{The $\chi^2$ significance of a dipole axis to be formed by the distribution of galaxy spin directions in different parts of the sky.} 
\label{dipole_all}
\end{figure*}

Figure~\ref{dipole_all} shows the $\frac{\chi^2_{real}-\chi^2_{random}}{\sigma(\chi^2_{\alpha,\delta})}$ for all possible $(\alpha,\delta)$ integer combinations. Code and data to reproduce the experiment is available at \url{https://people.cs.ksu.edu/~lshamir/data/asymmetry_hsc}.

The most likely dipole axis is at $(\alpha=94^o,\delta=52^o)$, with statistical significance of 3.3$\sigma$. The $1\sigma$ error range of the location is $(39^o,196^o)$ for the RA and $(-10^o,81^o)$ for the declination. HSC footprint that also overlaps with SDSS for the spectra is relatively small, at approximately 416 degree$^2$ \citep{more2023hyper}. The analysis identifies the best fit of a dipole axis by fitting all galaxies into a dipole axis alignment. Therefore, it is not affected by underpopulated parts of the sky that can skew the location of the most likely axis. But a smaller footprint also makes the location of the most likely dipole axis less accurate compare to previous studies that used far larger footprints.

\begin{figure*}[h]
\centering
\includegraphics[scale=0.25]{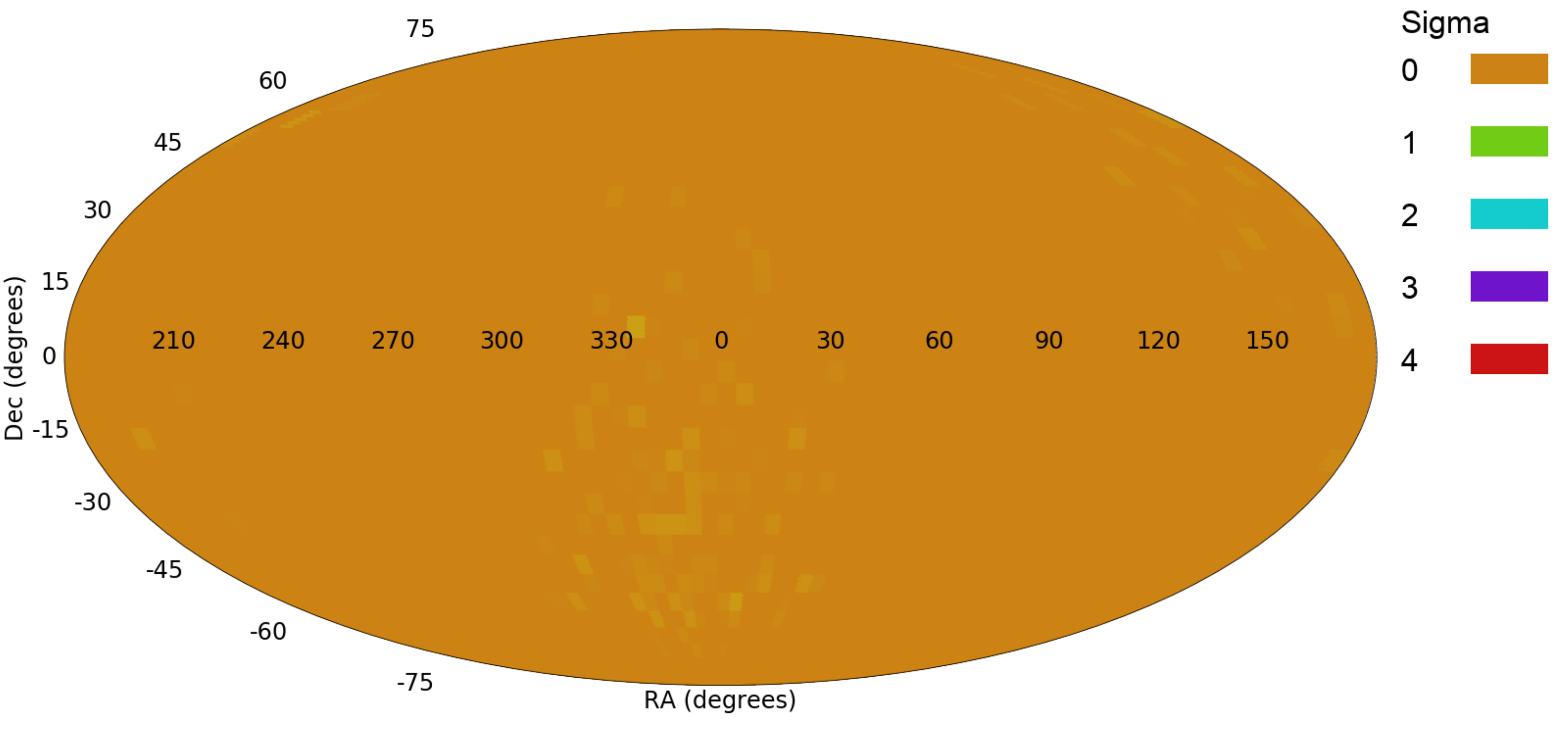}
\caption{The $\chi^2$ significance of a dipole axis to be formed in different parts of the sky when the galaxies are assigned with random spin directions.} 
\label{hsc_random_results}
\end{figure*}

While the relatively small footprint does not allow to identify the dipole axis with high accuracy, its location is still within the 1$\sigma$ error from the galactic pole at approximately $(\alpha=192^o,\delta=28^o)$. When assigning the galaxies with random spin directions instead of the observed spin directions, the maximum observed statistical significance drops to 0.51. That is expected as subtracting the $\chi^2$ of two sets of random values is not expected to generate signal. Reproducible analysis when assigning the galaxies with random directions of rotation is also available at \url{https://people.cs.ksu.edu/~lshamir/data/asymmetry_hsc}. Figure~\ref{hsc_random_results} shows the probability to have a dipole axis by chance in different parts of the sky when the galaxies are assigned with random directions of rotation.

\begin{figure*}[h!]
\centering
\includegraphics[scale=0.25]{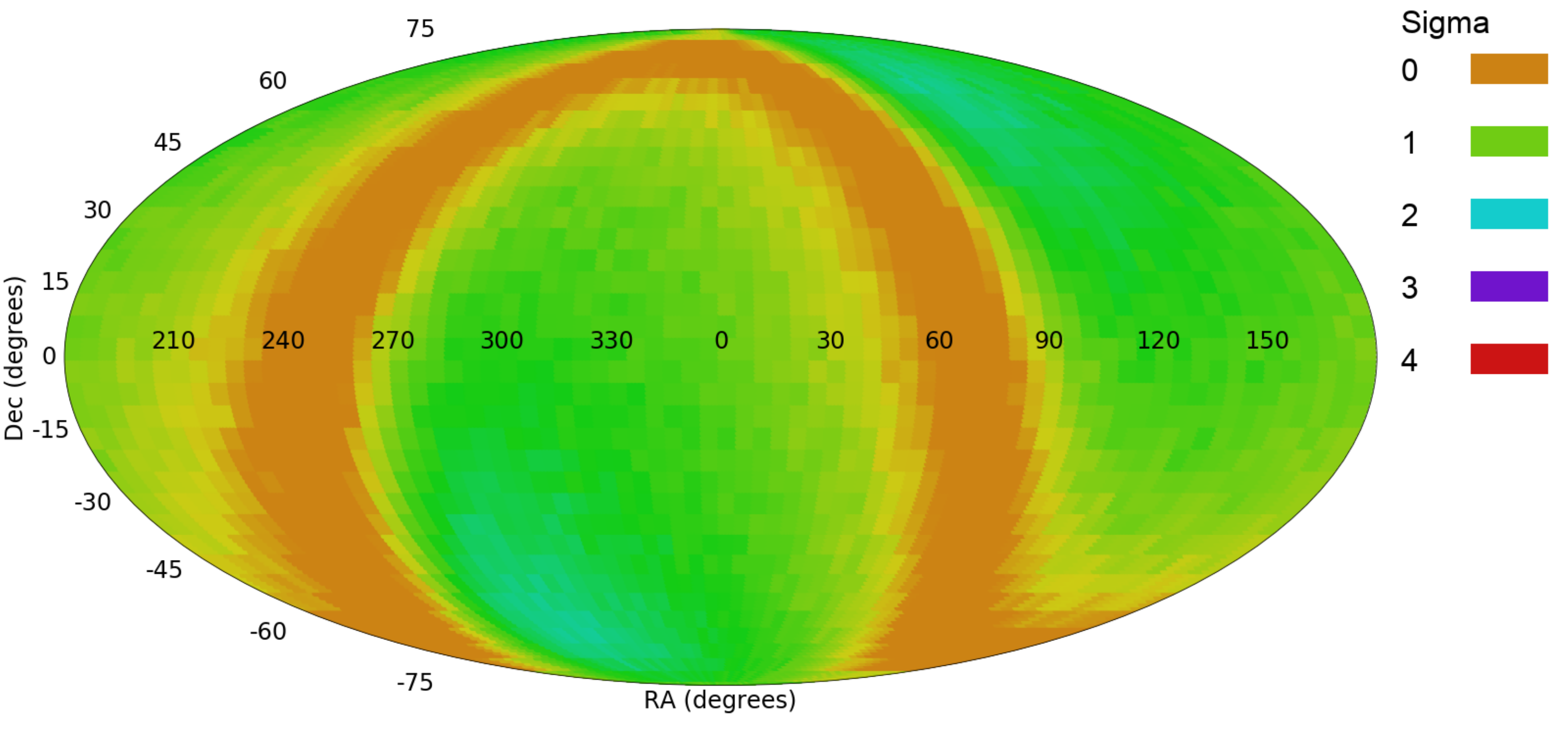}
\includegraphics[scale=0.25]{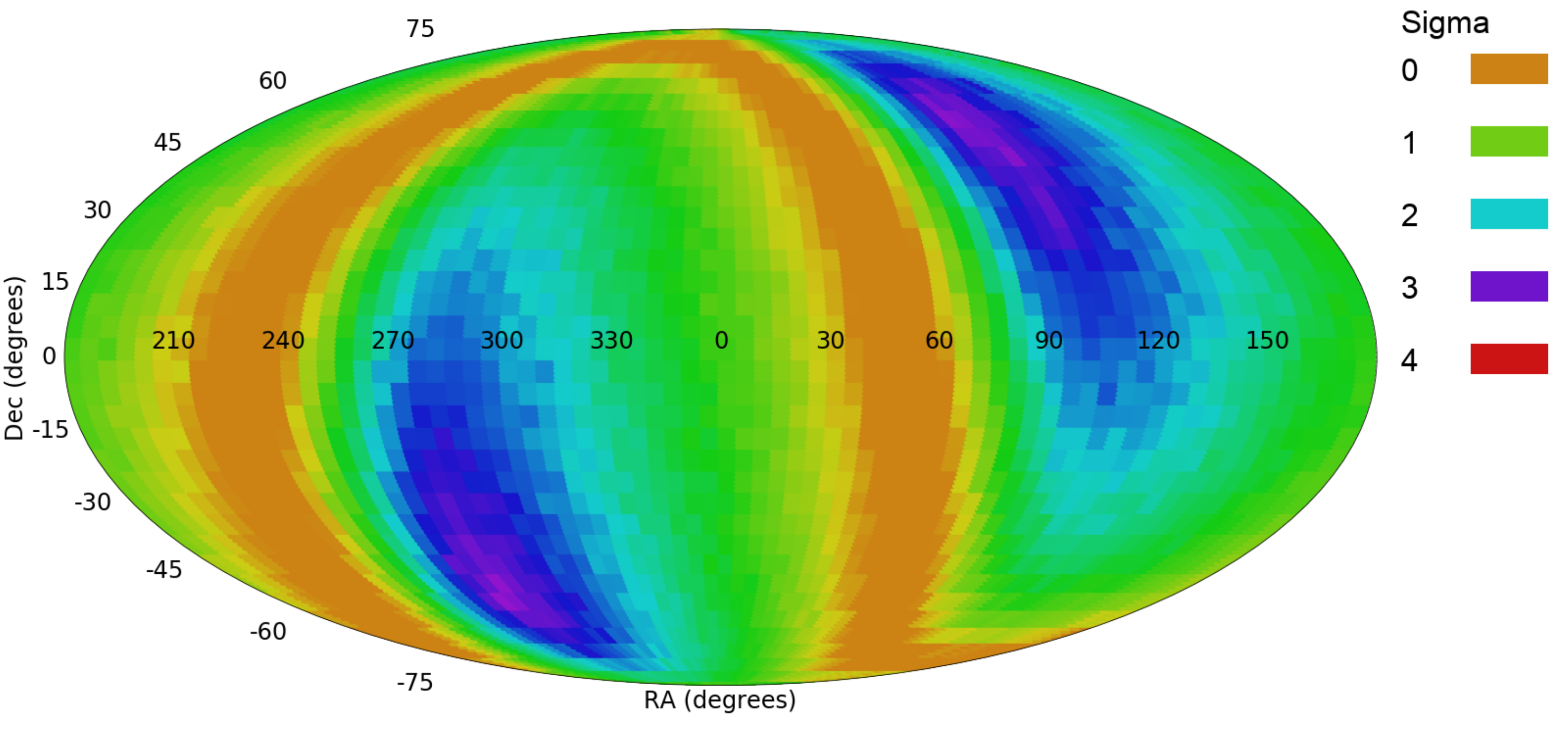}
\caption{The $\chi^2$ significance of a dipole axis to be formed by the distribution of galaxy spin directions in redshift of 0 to 0.1 (top) and in redshift of 0.1 to 0.2 (bottom). When using just galaxies with higher redshift range the statistical signal is stronger. The number of galaxies used in the two experiment is roughly the same, with 6,246 and 6,214 galaxies in the lower and higher redshift ranges, respectively. The two datasets are not overlapping, and no galaxy was used in both experiments}.
\label{dipole_redshift}
\end{figure*}

To test the link between the asymmetry and the redshift, the dataset can be divided into two redshift ranges: Galaxies of redshfit between 0 and 0.1, and galaxies with reshift of 0.1 to 0.2. The two sets of galaxies has 6,246, and 6,214 galaxies, respectively. That makes two sets of galaxies of similar sizes. Figure~\ref{dipole_redshift} shows the results of the analysis using each of these two redshift ranges, each range separately. 

When the redshift of the galaxies is $z<0.1$ the statistical strength of the axis is 1.79$\sigma$ peaking at $(\alpha=114^o,\delta=62)$. When the redshift range is $0.1<z<0.2$, the statistical strength of the dipole increases to 3.1$\sigma$, peaking at $(\alpha=91^o,\delta=50)$. Although the two experiments were based on two non-overlapping sets of galaxies, the location of the most likely axes are close to each other, and well within the 1$\sigma$ error of the dipole axis shown in Figure~\ref{dipole_all}. The stronger signal when the galaxies have a higher redshift is in agreement with previous observations using other telescopes, all of them show that the magnitude of the asymmetry increases as the redshift gets higher \citep{shamir2020patterns,shamir2022large}. When using high redshift galaxies imaged by JWST, the difference between the number of galaxies rotating in the same direction relative to the Milky Way and in the opposite direction relative to the Milky Way becomes extreme, and a statistically significant difference can be even observed by the naked eye, with no need for automatic annotation of large datasets of galaxies \citep{shamir2024galaxy}.

\section{Conclusion}
\label{conclusion}

While the directions of rotation of galaxies as observed from Earth is expected to be random, several experiments have suggested that the distribution is not necessarily random. These experiments are based on several different Earth-based and space-based telescopes and both the Northern and Southern hemispheres. This paper shows the analysis of the possible non-randomness using data provided by HSC. While HSC has a relatively small footprint, it is deeper than all other existing Earth-based digital sky surveys. That allows to profile the asymmetry in different redshift ranges. 

The results show that galaxies that rotate in the same direction relative to the Milky Way galaxy as observed from Earth are less prevalent compared to galaxies that rotate in the opposite direction relative to the Milky Way. It also shows that the signal becomes stronger when the redshift of the galaxy population gets higher. Explanations can be related to anomalies of the large-scale structure of the Universe, or to the physics of galaxy rotation.

If the galaxies had alignment in their inclinations or in their directions of rotation, asymmetry in the spin directions as observed from Earth could be expected. But since the inclination of galaxies is expected to be completely random, as well as the direction of rotation, the probability of a galaxy to rotate in a certain direction as observed from Earth is the same as its probability to rotate in the opposite direction. Therefore, asymmetry in the distribution of galaxy spin directions is not expected. But as also discussed in Section~\ref{results}, such asymmetry can be driven by differences in the brightness of galaxies, as galaxies that rotate in the same direction relative to the Milky Way are expected to be slightly brighter than galaxies that rotate in the same direction relative to the Milky Way.

The reproduction of scientific results has been a growing challenge in all fields of science, as most results published in scientific papers have been shown to be unreproducible \citep{stodden2018empirical}. That ``reproducibility crisis" naturally introduces a challenge for advancing science. For that reason, the results shown in this paper are fully reproducible, with code, data, and step-by-step instructions to easily reproduce the results. That allows the entire scientific community to ensure that the data used in this study is as described in the paper, and that the results shown in the paper using that data are correct. These results are also aligned with a large collection of observations that challenge the cosmological isotropy assumption \citep{aluri2023observable}, as well as other recent observations and tensions that challenge the standard cosmological model \citep{aluri2023observable,akarsu2024lambda,Timkov_2024,lopes2024dipolar}.

\section{Discussion}
\label{discussion}

Despite over a century of research, the physics of galaxy rotation is still one of the most provocative scientific phenomena, and research efforts to fully understand the nature of galaxy rotation are still being continued. The anomaly in the galaxy rotation was observed as early as the first half of the 20th century, proposing that the nature of galaxy rotation can be explained by the contention that the visible matter of the galaxy is embedded inside a much larger halo of highly dense non-luminous matter \citep{oort1940some}. These observations were ignored for nearly five decades, possibly due to their disagreement with the standard theories of the time \citep{rubin2000one}. Currently, the most common explanation to the anomaly of the galaxy rotation curve is the presence of dark matter \citep{rubin1983rotation}. Another common explanation is driven by certain modifications of Newtonian dynamics \citep{milgrom1983modification,bekenstein1984does,milgrom2009bimetric,falcon2021large}, and other explanations have also been proposed \citep{sanders1990mass,capozziello2012dark,chadwick2013gravitational,farnes2018unifying,rivera2020alternative,nagao2020galactic,sivaram2020mond,blake2021relativistic,gomel2021effects,skordis2021new,larin2022towards}.

Previous experiments provided evidence that the photometry and spectroscopy of galaxies observed from Earth are affected by the rotational velocity of these galaxies relative to the rotational velocity of the Milky Way. For instance, it has been shown that galaxies that rotate in the opposite direction relative to the Milky Way have lower redshift than galaxies spinning in the same direction relative to the Milky Way \citep{universe10030129,shamir2024empirical}. The data has also been shown that galaxies that rotate in the same direction as the Milky Way are dimmer than galaxies that spin in the opposite direction relative to the Milky Way \citep{shamir2020asymmetry,mcadam2023asymmetry}. Like this paper, these studies are provided with data, and the results are fully reproducible. While a small difference in the brightness is expected, the observed difference in brightness is larger than expected. That could be attributed to the yet unexplained physics of galaxy rotation, although there could be other explanations related to new physics or anomalies in the large-scale structure of the Universe \citep{shamir2020asymmetry,mcadam2023asymmetry}. 

If the brightness of galaxies as observed from Earth depends on the direction of rotation of the galaxies relative to the Milky Way galaxy, more galaxies that rotate in the opposite direction relative to the Milky Way will be more prevalent near both ends of the Galactic pole to an Earth-based observer. That is not because these galaxies are more prevalent in the Universe, but because they are brighter and therefore easier to detect from Earth. That agrees with the observation described here, showing an excessive number of galaxies that rotate in the opposite direction compared to the Milky Way.

On the other hand, large-scale alignment of the directions towards which galaxies spin has also been reported \citep{jones2010fossil,tempel2013evidence,tempel2013galaxy,codis2015spin,pahwa2016alignment,ganeshaiah2018cosmic,ganeshaiah2019cosmic,blue2020chiles,welker2020sami,kraljic2021sdss,lopez2021deviations}. If the cosmological-scale alignment forms a cosmological-scale axis that is not the Galactic pole, it can be considered a feature of the large-scale structure of the Universe. In that case, such a cosmological-scale axis can be aligned with alternative cosmological theories such as spinor-driven inflation \citep{bohmer2008cmb}, ellipsoidal universe \citep{campanelli2006ellipsoidal,campanelli2007cosmic,campanelli2011cosmic,gruppuso2007complete,cea2014ellipsoidal,tedesco2024ellipsoidal}, rotating universe \citep{godel1949example,ozsvath1962finite,ozsvath2001approaches,sivaram2012primordial,chechin2016rotation,seshavatharam2020integrated,camp2021}, black hole cosmology \citep{pathria1972universe,stuckey1994observable,easson2001universe,seshavatharam2010physics,poplawski2010radial,tatum2018flat,christillin2014machian,chakrabarty2020toy,poplawski2010cosmology,seshavatharam2014understanding,seshavatharam2020integrated,camp2021}, holographic universe \citep{susskind1995world,bak2000holographic,bousso2002holographic,myung2005holographic,hu2006interacting,rinaldi2022matrix,sivaram2013holography,shor2021representation}, and others \citep{astronomy3030014,ashmore2024data}.

If the asymmetry grows as the redshift gets higher as shown by the HSC data, it can provide an indication that the early Universe was more consistent than the present Universe, and gradually became more chaotic as reflected by the distribution of the directions towards which galaxies rotate. That can be aligned with cosmological models that are based on the contention that the Universe was born spinning, such as rotating universe \citep{godel1949example,ozsvath1962finite,ozsvath2001approaches,sivaram2012primordial,chechin2016rotation,seshavatharam2020integrated,camp2021} or black hole cosmology \citep{pathria1972universe,stuckey1994observable,easson2001universe,seshavatharam2010physics,poplawski2010radial,tatum2018flat,christillin2014machian,chakrabarty2020toy,poplawski2010cosmology,seshavatharam2014understanding,seshavatharam2020integrated,camp2021}.

But as described above, the results shown here can also indicate that the photometry of galaxies is affected by their rotational velocity relative to the rotational velocity of the Milky Way. In that case, the photometry has small but consistent bias as shown in \citep{mcadam2023asymmetry,shamir2017large,shamir2020asymmetry}, and a similar analysis has shown bias in the redshift \citep{shamir2024simple,shamir2024empirical}. A consistent bias in the redshift can be linked to the $H_o$ tension \citep{capozziello2024critical,vagnozzi2020new,rameez2021there,di2021realm,aloni2022step,davis2019can}, and possibly the tension between the size and shape of high-redshift galaxies observed with JWST and their expected age as determined by their redshift \citep{glazebrook2024massive}. A bias in the redshift, if such indeed exists, can explain this puzzling tension.



\section*{Data Availability}

Code, data, and step-by-step instructions to reproduce the results are available at \url{https://people.cs.ksu.edu/~lshamir/data/asymmetry_hsc/}.

\section*{Acknowledgments}
              
I would like to thank the two knowledgeable reviewers for the insightful comments. The research was funded in part by NSF grant 2148878.

\bibliographystyle{apalike}

\bibliography{main}

\end{document}